\documentclass[12pt]{article}
\input epsf
\newcommand{\ba}{\begin{array}}
\newcommand{\ea}{\end{array}}

\newcommand{\be}{\begin{equation}}
\newcommand{\ee}{\end{equation}}
\newcommand{\bea}{\begin{eqnarray}}
\newcommand{\eea}{\end{eqnarray}}
\newcommand{\beal}{\setcounter{letter}{1} \begin{eqnarray}}
\newcommand{\eeal}{\addtocounter{equation}{1} \end{eqnarray}}
\newcommand{\none}{\nonumber \\}
\newcommand{\req}[1]{Eq.(\ref{#1})}

\newcommand{\larrow}{\,\,\,\,\hbox to 30pt{\rightarrowfill}
\,\,\,\,}
\newcommand{\slarrow}{\,\,\,\hbox to 20pt{\rightarrowfill}
\,\,\,}

\newcommand{\half}{{1\over2}}

\begin{document}

\begin{titlepage}
\renewcommand{\thefootnote}{\fnsymbol{footnote}}
\renewcommand{\baselinestretch}{1.3}
\medskip
\hfill  UNB Technical Report 06-01\\[20pt]

\begin{center}
{\large {\bf The Fixed Points of RG Flow with a Tachyon  }}
\\ \medskip  {}
\medskip

\renewcommand{\baselinestretch}{1}
{\bf
J. Gegenberg $\dagger$\footnote{E-mail: geg@unb.ca},  
V. Suneeta $\sharp$\footnote{E-mail: suneeta@math.unb.ca} 
\\}
\vspace*{0.50cm}
{\sl
$\dagger$ Dept. of Mathematics and Statistics and Department of Physics\\
University of New Brunswick\\
Fredericton, New Brunswick, Canada  E3B 5A3\\ [5pt]
}
{\sl
$\sharp$ Dept. of Mathematics and Statistics\\
University of New Brunswick\\
Fredericton, New Brunswick, Canada  E3B 5A3\\ [5pt]
}

\end{center}

\renewcommand{\baselinestretch}{1}

{\bf Abstract}
We examine the fixed points to first-order  
RG flow of a non-linear sigma model with background metric, dilaton and 
tachyon fields. We show that on compact target spaces, the existence 
of fixed points with non-zero tachyon is linked to the sign of the 
second derivative of the tachyon potential $V''(T)$ (this is
the analogue of a result of Bourguignon for the zero-tachyon case). 
For a tachyon potential
with only the leading term, such fixed points are possible.
On non-compact target spaces, we introduce a small non-zero tachyon and
compute the correction to the Euclidean 2d black hole (cigar) solution
at second order in perturbation theory with a tachyon potential containing a 
cubic term as well. The corrections to the metric,
tachyon and dilaton are well-behaved at this order and tachyon `hair' 
persists.
We also 
briefly discuss solutions to the RG flow equations in the presence of 
a tachyon that suggest a comparison to dynamical fixed point 
solutions obtained by Yang and Zwiebach.
{\small
}\vfill
\hfill  May 2006 \\
\end{titlepage}

\section{Introduction}

Recently there has been extensive work on closed string tachyons and closed
string tachyon condensation. A review of this can be found in \cite{HMT}.
Some of the further papers on closed string tachyons in various contexts
have helped to better understand the effects of localized, winding and bulk
tachyons \cite{after2004}. Also, Yang and Zwiebach \cite{YZ} have
investigated a class of fixed-point solutions of the first order tachyon-dilaton-metric RG flow
for the non-linear sigma model of bosonic closed string theory.
 
In this paper, we investigate the question of existence of non-trivial fixed points of the 
first order RG flow equations for the metric-tachyon-dilaton system 
in a more general context on compact and non-compact target spaces. We are interested primarily
in addressing questions such as the following: Is it possible to link the question of existence of
solutions with non-zero tachyon for these equations to {\em the form of the 
tachyon potential}, and is there an analogue of the 2d Witten black hole type
solution  in the presence of a tachyon? The first question can be answered
in the affirmative on compact target spaces, and we can only deal with the
second question perturbatively (on non-compact target spaces). 

The fixed point equations for the metric and tachyon are
\begin{eqnarray}
R_{ij} + 2 \nabla_{i} \nabla_{j} \Phi - ( \nabla_{i} T)( \nabla_{j} T) = 0,
\label{eq1-1}
\end{eqnarray}
\begin{eqnarray}
\Delta T - V'(T) - 2 ( \nabla_{i} \Phi) (\nabla^{i} T) = 0.
\label{eq1-2}
\end{eqnarray}
Here, $T(x)$ is the tachyon, $\Phi (x)$ the dilaton and $V(T)$ is the
tachyon potential whose derivative appears in Eq.(\ref{eq1-2}).

These two equations above (i.e. beta functions of the metric and tachyon 
being set to zero) imply that the dilaton
beta function is a constant; i.e
\begin{equation}
- \frac{1}{2} \left ( \Delta \Phi - 2 |\partial_i \Phi|^2 - V(T) \right ) = c.
\label{eq1-3}
\end{equation}

To see this, we start by taking the covariant divergence of \req{eq1-1}.  Then we use the 
contracted Bianchi identity $\nabla_j R_i^j= \frac{1}{2} \nabla_i R$ and the contracted Ricci identity in 
the form
$$
\nabla_j\nabla^j\nabla_i\phi=\nabla_i(\Delta\phi)+R_{ij}\nabla^j\phi.
$$
After using $\nabla^j T \nabla_i(\nabla_j T)=\half\nabla_i(|\nabla T|^2)$, 
$\nabla_i T V'(T)=\nabla_i (V(T))$ and \req{eq1-2} we 
get
$$
\nabla_i\left(\half R+2\Delta\phi-\half|\nabla T|^2-V(T)\right)+2(R_{ij}-\nabla_i T\nabla_j T)\nabla^j\phi=0.
$$
Now, using \req{eq1-1} and its contraction we get
$$
\nabla_i\left(\Delta\phi-2|\nabla\phi|^2-V(T)\right)=0.
$$
We get the desired result upon noting that the term in brackets above is twice the 
usual expression for the beta function of the dilaton $\phi$.

These solutions are the analogues of similar results obtained for
other string theory RG flows, for example in \cite{cfmp, curci, osborn}.
As is well known, $c$ is then the central charge of the resulting
sigma model, which is a CFT. When $c=0$, some properties of solutions to 
Eq.(\ref{eq1-1} - \ref{eq1-3}) were discussed extensively by 
Yang and Zwiebach \cite{YZ}.  

We will discuss solutions to Eq.(\ref{eq1-1} - \ref{eq1-3}) for any value of $c$. 
These solutions are analogous, for example, to the Witten 2d black hole (cigar)
solution when $T=0$ \cite{witten, wadia}. In fact, when $T=0$, there are many results that
restrict the non-Ricci-flat solutions to Eq.(\ref{eq1-1} - \ref{eq1-3}) 
\footnote{ The case $T=$ (non-zero) constant differs from $T=0$ in the analysis of this
paper only by a
shifted central charge, the shift being the constant tachyon potential; so
we do not consider it separately.}. 
On non-compact target spaces, the only 
such known solution is the cigar; in fact, it is now known that
any non-Ricci flat solution must either have the integral of its scalar
curvature unbounded, or the diffeomorphism generated by the dilaton must
violate a certain asymptotic condition, the details of which are 
discussed in \cite{osw}. On {\em compact spaces}, when $T=0$, there are no
solutions apart from the Ricci-flat metrics, by a result due to
J-P Bourguignon \cite{JPB}. 

We are interested in non-trivial solutions 
to these equations with Euclidean signature metrics - we define non-trivial
solutions to mean those for which $T \neq 0$ (and non-constant).
In section 2, we discuss compact target spaces. We make no 
assumptions in this section on the form of $V(T)$; rather, we would like
to find out if it is possible to obtain a general result for $T \neq 0$
(similar to the $T=0$ result in \cite{JPB}) on (non)-existence of solutions. The hope is that the 
the existence or non-existence of solutions is somehow linked to the
form of $V(T)$. This hope is indeed realised; the existence of solutions is
linked to the sign of $V''(T)$. In particular, if $V''(T) < 0$ everywhere on
the target space (which happens if we only consider the leading term in the
tachyon potential), then {\em non-trivial solutions cannot be ruled out}. 
This is in contrast to the $T=0$ case, which we can see by setting $T=0$ in the 
proof in section 2. In fact, we then reproduce Bourguignon's proof \cite{JPB}. 
When $T \neq 0$, the conditions on $V''(T)$ are actually even more
general, and are summarised at the end of section 2 in the form of two cases.

In section 3, we consider non-compact target spaces. Here it is difficult
to obtain general results, so instead, we ask if there is an analogue of the 
2d Witten black hole (cigar) solution 
in the presence of a non-zero tachyon. We are only
able to address the problem perturbatively - and for a particular $V(T)$ 
that includes terms up to cubic order in the tachyon. 
The problem has already been discussed perturbatively from different points
of view in \cite{witten, wadia}. In \cite{marcus, peet}, it was found that
with quadratic tachyon potentials, there was an exact solution to the tachyon 
back-reaction which was regular at the horizon, but did not fall off
fast enough asymptotically - leading to tachyon `hair'.
We investigate what happens to `hair' when considering a tachyon potential
with a cubic term as well.  
We introduce a perturbation
parameter (which is just the magnitude of the tachyon we have introduced),
with respect to which we find a second-order correction to the tachyon.
Both the metric and tachyonic perturbations
are bounded at this order, and the asymptotic properties of the metric, dilaton and tachyon do not change substantially - in fact the corrections 
to the tachyon resulting from the cubic term fall off much faster 
asymptotically. Thus for tachyons of small magnitude, there seems to be no
serious problem at this stage in perturbation theory, and tachyon
hair persists. Whether this 
perturbative solution can actually be continued to all orders in perturbation 
theory remains an open question.

Finally, in the last section, we write down the RG {\em flow} equations in the presence of
the tachyon. We 
compute solutions to these flow equations for two choices of the tachyon potential. When the 
tachyon potential has only the leading term, the solutions obtained closely
resemble dynamical fixed-point spacetime solutions of Yang and Zwiebach.
However this holds only for early times if the potential is modified.  
These two examples illustrate that the RG flow parameter does not always
play the role of dynamical time.

\section{Fixed points with a non-zero tachyon on compact target spaces}

We obtain results on solutions to Eq.(\ref{eq1-1}, \ref{eq1-2})  on compact target spaces with Euclidean signature
metrics. As we saw, it follows that Eq.(\ref{eq1-3}) is automatically satisfied for some $c$.
We define $v_{i} = - \nabla_{i} \Phi $ and rewrite equations Eq.(\ref{eq1-1}, \ref{eq1-2})  as
\begin{eqnarray}
R_{ij} = \nabla_{i} v_{j} + \nabla_{j} v_{i} + ( \partial_{i} T)( \partial_{j} T),
\label{eq2-1}
\end{eqnarray}
\begin{eqnarray}
\Delta T - V'(T) + 2 v_{i} (\nabla^{i} T) = 0.
\label{eq2-2}
\end{eqnarray}

We would ideally like to link existence of solutions to the above equations
to the form of $V(T)$; so we will make no assumptions about its form at 
this stage. We demonstrate in the rest of the section how this is realised.
The main objective is to derive Eq.(\ref{eq2-15}) - the 
conditions for existence of solutions then follow from an analysis of this
equation. This analysis is done by considering two cases for the 
behaviour of $V(T)$, both of which, it seems, cannot be ruled out for 
tachyonic potentials. However, if $V(T)$ contains only the leading order
term, then it falls into what we refer to at the end of this section as case 2.

We first take the double divergence of Eq.(\ref{eq2-1}) and use Bianchi identities.
We then get
\begin{eqnarray}
\frac{1}{2} \Delta R = \nabla^{j}(\Delta v_{j}) +
\nabla^{j} \nabla^{i} (\nabla_{j} v_{i}) )  
+ \left [ \nabla^{j}(\Delta T)
\right ] (\nabla_{j} T) + (\Delta T)^{2} \nonumber \\
+ (\nabla^{i} \nabla^{j} T)(\nabla_{i} \nabla_{j} T) 
+ (\nabla_{i} T)(\nabla^{j}\nabla^{i}\nabla_{j} T).
\label{eq2-3}
\end{eqnarray}

To simplify the above expression, we note upon using the Bianchi and Ricci
identities, that

\begin{eqnarray}
\nabla^{j} \nabla^{i} \nabla_{j} v_{i}
= \Delta(\nabla \cdot v) + R_{~j}^{m} (\nabla^{j}v_{m}) + \frac{1}{2}
(\nabla^{m} R)v_{m} \nonumber \\
= \nabla^{j}(\Delta v_{j}).
\label{eq2-4}
\end{eqnarray}
Further,
\begin{eqnarray}
(\nabla_{i} T)(\nabla^{j}\nabla^{i}\nabla_{j}T)
= (\nabla_{i}T) \nabla^{i}(\Delta T) + R^{mi}(\nabla_{i}T)(\nabla_{m}T).
\label{eq2-5}
\end{eqnarray}

Taking the trace of Eq.(\ref{eq2-1}) gives
\begin{eqnarray}
\nabla \cdot v = \frac{1}{2} (R - \vert \nabla_{i} T \vert^2 ).
\label{eq2-6}
\end{eqnarray}

Substituting the results of Eq.(\ref{eq2-4}, \ref{eq2-5}, \ref{eq2-6}) into 
Eq.(\ref{eq2-3}) and using Eq.(\ref{eq2-1}),
we get
\begin{eqnarray}
\frac{1}{2} \Delta R - \Delta (\vert \nabla_{i} T \vert^2 ) + (\nabla^m R)v_m 
+ R^{lm} \left [ R_{lm} - (\nabla_{l}T)(\nabla_{m}T) \right ] \nonumber \\
+ \left [ \nabla^{j}(\Delta T) \right ](\nabla_{j}T) + (\Delta T)^2 + \vert (\nabla^{j}\nabla^{i}T) \vert^2 + (\nabla_{i}T)\nabla^{i}(\Delta T) \nonumber \\
+ R^{im} (\nabla_{i}T)(\nabla_{m}T) = 0.
\label{eq2-7}
\end{eqnarray}

Now, we note that (Bochner formula)
\begin{eqnarray}
\Delta (\vert \nabla T \vert^{2})
= 2 \left [ \Delta (\nabla_{i}T)\right ](\nabla^{i}T) +
2 \vert (\nabla_{i}\nabla_{j}T) \vert^2 \nonumber \\
= 2 (\nabla^{i}T)\nabla_{i}(\Delta T) + 2 R^{ij}(\nabla_{i}T)(\nabla_{j}T)
+ 2 \vert (\nabla_{i}\nabla_{j}T) \vert^2.
\label{eq2-8}
\end{eqnarray}

Substituting this back into Eq.(\ref{eq2-7}) and simplifying,
we get

\begin{eqnarray}
\frac{1}{2} \Delta R - 2 R^{ij}(\nabla_{i}T)(\nabla_{j}T)
- \vert (\nabla_{i}\nabla_{j}T) \vert^2 
+  (\nabla^m R)v_m + \vert R_{ij} \vert^2 \nonumber \\
+ (\Delta T)^2 = 0.
\label{eq2-9}
\end{eqnarray}

Now, adding and subtracting terms, we can rewrite
\begin{eqnarray}
(\nabla^m R)v_m = \left [ \nabla^m ( R - \vert \nabla_{i} T \vert^2 ) \right ]v_m
+ (\nabla^m \vert \nabla_{i} T \vert^2 )v_m,
\label{eq2-10}
\end{eqnarray}
and then do the following manipulations:
\begin{eqnarray}
(\nabla^m \vert \nabla T \vert^2 )v_m
= 2 v_m \left [ \nabla^{m}\nabla^{i}T \right ](\nabla_{i}T) \nonumber \\
= 2 v_m \left [ \nabla^{i}\nabla^{m}T \right ](\nabla_{i}T) \nonumber \\
= 2 \left [ \nabla^{i} (v^{m}\nabla_{m}T) \right ](\nabla_{i}T)
- 2 (\nabla^{i} v^{m})(\nabla_m T)(\nabla_i T) \nonumber \\
=  \left [ \nabla^{i} (2 v^{m}\nabla_{m}T) \right ](\nabla_{i}T)
- \left [ R^{im} - (\nabla^i T)(\nabla^m T) \right ] (\nabla_i T)(\nabla_m T).
\label{eq2-11}
\end{eqnarray}

Now, we use the fixed-point condition for the tachyon, Eq.(\ref{eq2-2}) and
substitute for $2 v^{m}\nabla_{m}T$.
We finally get
\begin{eqnarray}
(\nabla^m R)v_m = \nabla^m \left [ ~R - \vert \nabla_i T \vert^2 ~ \right ]v_m 
+ V''(T) (\vert \nabla_{i} T \vert^2 ) - [ \nabla^{i}(\Delta T) ](\nabla_i T) \nonumber \\
- R^{ij} (\nabla_{i}T)(\nabla_{j}T) 
+
\left ( \vert \nabla_{i}T \vert^2 \right )^2 .
\label{eq2-12}
\end{eqnarray}
In the above equation, we used
$\left [ \nabla_i V'(T) \right ] (\nabla^i T) = V''(T) (\vert \nabla_{i} T \vert^2 )$.

We now substitute the result of Eq.(\ref{eq2-12}) into Eq.(\ref{eq2-9}) to obtain
\begin{eqnarray}
\frac{1}{2} \Delta R - 2 R^{ij}(\nabla_{i}T)(\nabla_{j}T)
- \vert (\nabla_{i} \nabla_{j}T) \vert^2 
+ \vert R_{ij} \vert^2 + (\Delta T)^2  \nonumber \\
+ \nabla^m \left (~R - \vert \nabla_i T \vert^2 \right )v_m 
+ V''(T) (\vert \nabla_{i} T \vert^2 ) - [ \nabla^{i}(\Delta T) ](\nabla_i T) \nonumber \\
- R^{ij} (\nabla_{i}T)(\nabla_{j}T) 
+
\left ( \vert \nabla_{i}T \vert^2 \right )^2
= 0.
\label{eq2-13}
\end{eqnarray}

Now we add and subtract on the left hand side of Eq.(\ref{eq2-13}),
the term \\
$1/2 \Delta(\vert \nabla_{i} T \vert^2 )$. Eq.(\ref{eq2-8}) implies
\begin{eqnarray}
\frac{1}{2} \Delta(\vert \nabla_{i} T \vert^2 ) =
\vert \nabla_i \nabla_j T \vert^2 + R^{ij} (\nabla_i T)(\nabla_j T)
+ (\nabla_i T)[ \nabla^{i}(\Delta T)].
\label{eq2-14}
\end{eqnarray}

Finally, after some manipulations, we get
\begin{eqnarray}
\Delta[R - \vert \nabla_i T \vert^2 ] =
-2 \nabla^m [R - \vert \nabla_i T \vert^2 ]v_m
-2 \vert R_{ij} - (\nabla_i T)(\nabla_j T) \vert^2 \nonumber \\
- 2(\Delta T)^2 
-2 V''(T) \vert \nabla_i T \vert^2.
\label{eq2-15}
\end{eqnarray}

As we said at the beginning of this section, obtaining the above equation was
the primary objective. As we see below, it leads to statements on existence 
of solutions to Eq.(\ref{eq2-1}, \ref{eq2-2}). Also, the result of
Bourguignon \cite{JPB} is obtained by setting $T=0$ in the argument below.

In what follows, we assume that $V'(T)$ is not a constant unless T=0 (or constant), in which case
$V'(T) = 0$. Essentially we are not considering linear potentials as they
are not relevant for a study of tachyons.

Denoting the target space by $M$, let us consider a point $p \in M$ where
$(R - \vert \nabla_i T \vert^2 )$ takes its global minimum. Recall that
since $M$ is compact, functions on $M$ either have a global minimum or
are constant throughout (in which case we can take any point $p \in M$). Then, at
this point
we have  $\nabla^m [R - \vert \nabla_i T \vert^2 ]_p = 0$ and
$ \Delta[R - \vert \nabla_i T \vert^2 ]_p \geq 0$. 

We discuss the 
following two cases, which are essentially conditions on $V(T)$ at 
the mimimum $p$.
Since these conditions are only required at one point in the target space (i.e
the minimum), it seems possible to have tachyonic potentials satisfying 
either of the cases. If, however, $V(T)$ had only the usual leading term, it
would satisfy case 2.

{\bf Case 1:} \\
We assume that at the global minimum  $p \in M$ of $(R - \vert \nabla_i T \vert^2 )$,
$V''(T) \geq 0$. 

Under this assumption, we now argue that no non-trivial steady solitons are
possible.

{\bf Proof: }\\
In this case, we have, at the minimum $p$, that the left-hand side of
Eq.(\ref{eq2-15}) is non-negative; the right-hand side is non-positive.
Thus the only possibility is that both sides are
exactly zero at the minimum. This is very restrictive because it
implies that each term in the right-hand side must vanish at $p$, so in
particular, $ R_{ij} - (\nabla_i T)(\nabla_j T) = 0$ at the minimum.
Taking the trace, we get that $[R - \vert \nabla_i T \vert^2 ]_p = 0$.
Therefore, everywhere else on the manifold $M$, we must have
$[R - \vert \nabla_i T \vert^2 ] \geq 0$. However, we can now integrate
both sides of Eq.(\ref{eq2-6}) and since we are on a compact manifold,
the integral of a total divergence is zero. So we get
\begin{equation}
\int_{M} [R - \vert \nabla_i T \vert^2 ]dV = 0.
\label{eq2-16}
\end{equation}

Since the integrand is non-negative, this equation implies that it
must be zero everywhere. So we have
$[R - \vert \nabla_i T \vert^2 ] = 0$ everywhere. Substituting this back in
Eq.(\ref{eq2-15}), the following equations then hold everywhere:
\begin{eqnarray}
R_{ij} - (\nabla_i T)(\nabla_j T) = 0; \nonumber \\
\Delta T = 0; \nonumber \\
V''(T) \vert \nabla_i T \vert^2 = 0.
\label{eq2-17}
\end{eqnarray}

Since we assume that $V'(T)$ is not a constant, 
the above equations imply that $T=0$ (or non-zero constant, see footnote on page 2). Thus we have shown only $T=0$ 
solutions are possible. However, when $T=0$, by Bourguignon's result \cite{JPB},
there are no solutions to the fixed point equations on compact manifolds other
than Ricci-flat metrics.
We see this by setting $T=0$ in Eq.(\ref{eq2-17}). Then $R_{ij} = 0$. 

{\bf Case 2:} \\
We assume that at the global minimum $p \in M$ of $\left(R-|\nabla_i T|^2\right)$, $V''(T) < 0$.
This is certainly possible. For example, taking a tachyon potential
with only the usual quadratic term, i.e $V(T) = - m^2 T^2$, we will
obtain $V''(T) < 0 $ everywhere, so we satisfy this condition {\em (although
we only want the condition to hold at $p$)}.
In this case, we note that we can again study Eq.(\ref{eq2-15}) at a global
minimum of  $[R - \vert \nabla_i T \vert^2 ]$. However, the right-hand side
of this equation is no longer non-positive, due to the sign of
$V''(T)$. So we cannot obtain conclusive results as in the previous case.
Thus, interestingly, we are left with the open possibility of 
fixed point solutions to the RG flow equations 
with a non-zero tachyon.

Thus we can conclude that Case 1 presents an obstruction to the 
existence of solutions to the first-order fixed point equations, whereas 
there is no such obstruction in Case 2, so solutions are allowed. 
It is implicitly assumed that the 
manifold is smooth, but we make no assumptions on the dilaton.
It is quite possible that if a solution existed in Case 2, it could 
either have high-curvature regions (necessitating higher $\alpha'$ 
corrections) or a dilaton such that the string coupling constant becomes
significant. Since we do not have an explicit solution, we are unable to
comment on whether either of these happen.

\section{Fixed points on non-compact target spaces: Perturbing the cigar}
In this section, we examine fixed points of Eqs.(\ref{eq1-1}, \ref{eq1-2}) on
non-compact target spaces. We are interested in non-Ricci-flat solutions with
$T \neq 0$.\footnote{The arguments in the previous section crucially
used compactness of the target space, and therefore some of the 
conclusions no longer apply.} 
When the tachyon is zero, there is a non-Ricci-flat solution to these equations
where the metric changes only by diffeomorphisms generated by the gradient
of the dilaton. 
This is the cigar solution, or Euclidean Witten black hole \cite{witten, 
wadia} since we only consider Riemannian metrics throughout this
paper.
The metric has the form
\be
ds^2=(1+r^2)^{-1}(dr^2+r^2 d\theta^2),\label{cigards}
\ee
while the dilaton is given by 
\be
\phi(r)=-\half\ln{(1+r^2)}.\label{cigardil}
\ee  

When there is a non-zero tachyon, the task of finding solutions is complicated.
In particular, whether there is an  analogue of the cigar solution in the presence of a tachyon
was attacked perturbatively in the first papers discussing the cigar solution,
i.e  
\cite{witten,wadia}. 
Subsequent work on this subject led to a discussion of the right boundary 
conditions to be employed at the horizon while studying tachyon perturbations.
With the boundary conditions demanding regularity at the horizon, tachyon
perturbations for quadratic tachyon potentials 
were studied in \cite{marcus, peet}. Exact solutions for the back-reaction 
of the tachyon were obtained in both these papers 
and the back-reacted metric and
dilaton were derived in \cite{marcus}. Very interestingly, the results of
\cite{marcus,peet} signal the presence of tachyon `hair' which both groups of
authors discuss extensively in the Euclidean and Lorentzian cases.

As an extension to the results of 
our previous section (which are valid for any general tachyon potential), 
we could ask what happens to tachyon hair in the 
presence of any general tachyon potential $V(T)$. Such a question is difficult 
to tackle in full generality on a non-compact target space, so 
we will take a tachyon potential with a quadratic as well as cubic term:
\be
V(T)=-\frac{m^2}{2}T^2+\beta T^3.\label{tachpot}
\ee
In the above, $m$ is the tachyon mass and $\beta$ is a constant of order 
unity \footnote{For example, in  Yang and Zweibach \cite{YZ} it is claimed that 
the closed 
string tachyon potential is given by \req{tachpot} with the ratio of 
$\beta$ to $m^2=1$ being  
$6561/4096$, up to terms of order $T^4$.}. This analysis is instructive and
the tachyon hair found in \cite{marcus,peet} persists at least for small
tachyon perturbations - in fact, corrections to the tachyonic perturbations
due to the cubic term in the potential fall off much faster asymptotically.
It seems likely that tachyon hair persists for any power law 
tachyon potential and there are no significant effects asymptotically
of the higher powers of $T$ in the potential.

We introduce
a small tachyonic perturbation and solve Eq.(\ref{eq1-1}, \ref{eq1-2})
in perturbation theory around the cigar solution - we would like to 
obtain the corrected metric, tachyon and dilaton with a tachyonic potential
of the form \req{tachpot}. The perturbation parameter is related to
the small amplitude of the tachyon we have introduced.  

We choose the circularly symmetric metric in the conformal gauge:
\be
ds^2=f(r)\left(dr^2+r^2d\theta^2\right),\label{axial}
\ee
and the tachyon $T=T(r)$. 

The fixed points of the flow satisfy \req{eq1-1} and \req{eq1-2}.
Substituting this {\it ansatz} into the $rr$ and $\theta\theta$ components of 
\req{eq1-1} and the 
tachyon equation \req{eq1-2}, respectively, 
we get
\bea
r f f''-r (f')^2+f f' + 2 r f(-2 f \phi''+f' \phi')+2 r f^2 (T')^2=0,
\label{rr}\\
r f f''-r (f')^2+f f'-2 f\phi'(r f'+2f)=0,\label{thetatheta}\\
r T''+\left(1-2 r\phi'\right) T'+ r f(m^2 T-3\beta T^2)=0.\label{T}
\eea

We know that if $T(r)=0$ everywhere, then we get the cigar 
solution \req{cigards} and \req{cigardil}\cite{witten,wadia}:

Hence we perturb around the cigar solution. We introduce a perturbation
parameter $0 \leq \epsilon < 1$, which is the magnitude of the tachyon field.  
Hence the perturbations of the metric and dilaton must start at order $\epsilon^2$ 
in this scheme, as discussed in \cite{peet}, in order to isolate the
effect of a small tachyon from other potential deformations.  
To wit:
\bea
T(r)&=&\sum_{a=1}\epsilon^a T_a(r),\\
f(r)&=&(1+r^2)^{-1}+\sum_{a=2}\epsilon^a f_a(r),\\
\phi(r)&=&-\half\ln(1+r^2)
+\sum_{a=2}\epsilon^a \phi_a(r).
\eea
To lowest order, the tachyon equation \req{T} is \cite{marcus,peet}
\be
r(1+r^2)T_1''+(1+3 r^2)T_1'+m^2 r T_1=0.\label{T1}
\ee
This is Legendre's equation if we change variable $r\to 1+2 r^2$. Two 
linearly independent solution in the region $r\geq 0$ are the Legendre 
functions $P_\nu(1+2 r^2)$ and $Q_\nu(1+2 r^2)$
of order $\nu:=-\frac{1}{2}\left(1+\sqrt{1-m^2}
\right)$.  The solution regular at the tip and in the asymptotic region 
is $P_\nu(1+2 r^2)$.  Another way to represent this solution is as 
the hypergeometric function ${}_2F_1(\half,\half,1,1-z)$ \cite{AS}, and 
it is this form of the solution that is used in the tachyon hair literature 
\cite{marcus, peet}.  

To order $\epsilon^2$, the metric and dilaton fields can be extracted from 
the explicit solutions displayed in the Appendix of \cite{marcus}. 
To this order, there is no contribution to these fields 
from the $T^3$ term in the tachyon potential.  However, to this order 
the tachyon (which has a contribution from this term
in the potential) satisfies \req{T} to order $\epsilon^2$:
\be
r(1+r^2)T_2''+(1+3 r^2)T_2'+m^2 r T_2=3\beta r P^2_\nu(1+2 r^2).
\ee
This equation has the form of the inhomogenized version of the homogeneous 
equation \req{T1}.  Hence the general solution for the second order 
perturbation 
$T_2(r)$ is a linear combination of Legendre functions of order $\nu$ with a 
particular solution.  The latter is easy to find, using the identity:
\be
\partial_r P_\nu(1+2 r^2) Q_\nu(1+2 r^2)-\partial_r Q_\nu(1+2 r^2) P_\nu(1+2 r^2)=\frac{1}{r(1+r^2)},
\ee
for Legendre functions of arbitrary order $\nu$ with the result that:
\bea
T_2(r)&=&3\beta P_\nu(1+2 r^2)\int  P_\nu^2(1+2 r^2) Q_\nu(1+2 r^2) r dr\\\none 
&-&3\beta Q_\nu(1+2 r^2)\int  P_\nu^3(1+2 r^2) r dr
.
\eea 
This perturbation is regular everywhere, and approaches the $O(\epsilon^1)$ approximation for 
$r\sim 0$ and for $r\to\infty$.  
In Figure 1, we show the MAPLE plots of $T_1(r)$ and $T_1(r)+ \epsilon T_2(r)$ with $\epsilon=0.1$ and 
$m=1$, that is, $\nu=-\half$.

We can study the asymptotic properties of the next-order perturbation $T_2 (r)$;
it falls off approximately two orders faster as a power of $r$ as compared to
$T_1$. Thus at order $\epsilon^2$, tachyon hair persists, and adding the 
correction $T_2$  
makes no difference asymptotically (as seen in Figure 1). \footnote{ In addition to the 
faster fall-off, $T_2$ is multiplied by one 
higher power of $\epsilon$ as compared to $T_1$ making its contribution
even smaller.}
Most probably, considering a tachyon potential with quartic or higher terms
will also cause no substantial changes to the asymptotic behaviour, 
and therefore not alter the presence of hair.
Further, the effect of the cubic term in the tachyon 
potential affects the metric and dilaton at order $\epsilon^3$, and we do not
expect a significant change asymptotically. 
We note that
in the limit as $m\to 0$ the perturbations blow up.  This reflects the fact that the solution for a 
`massless tachyon'- that is, just a massless scalar field, has singularities at $r=0$ and as $r\to\infty$.  

In this section, we have discussed perturbative solutions for small tachyon 
amplitudes. There is still the question of whether there is an exact 
solution to the first-order equations of motion. This certainly looks 
plausible since the corrections obtained at order
 $\epsilon^2$
are well-behaved at the horizon, and fall off asymptotically two
orders in $r$ faster than the 
lower order correction.

Is there an exact solution with a tachyon to all orders in $\alpha'$?
In \cite{witten}, it is shown that the sigma model with a cigar metric (i.e
in the absence of a tachyon) is
a CFT. 
Further, a `deformed cigar' solution was 
obtained in \cite{DVV}, exact to all orders in $\alpha'$ and solving
the Weyl invariance conditions of the metric {\em and} the dilaton.
Therefore, it seems reasonable that
there could be such a generalisation with a non-zero tachyon, \footnote{We note that \cite{DVV} mention an on-shell tachyon mode in their section on the deformed cigar.} - our 
analysis does not unfortunately permit us to comment any further.
There is, lastly, the issue of string loop corrections that become important
if the string coupling constant becomes strong anywhere. 
At least up to second order in $\epsilon$, the dilaton still approaches the
cigar dilaton asymptotically. In Schwarzschild-like coordinates, this is the
linear dilaton, and the coupling constant vanishes asymptotically. The dilaton
does not diverge at the horizon or anywhere in the interior in our Euclidean
analysis, so we conclude that string loop effects are not likely to produce 
a dramatic change to our discussion.

\clearpage
\begin{figure}[hbt]
\vspace{5cm}
\begin{center}
\leavevmode
\epsfxsize= 8 cm
\epsfbox{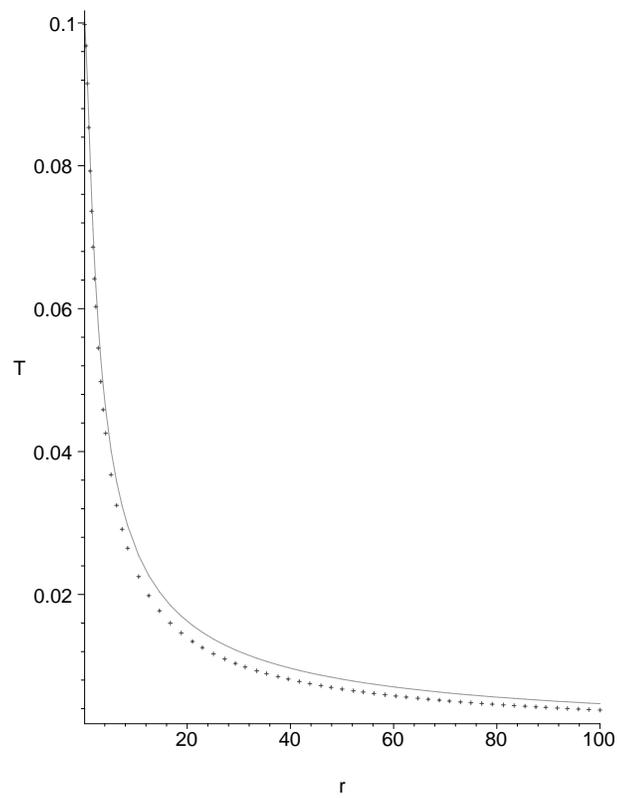}
\end{center}
\caption{Graph of $T(r)$.  The graph with the dotted line is that of $T_1$; the 
solid line is the graph of $T_1 + \epsilon T_2$.}
\label{Fig1}
\end{figure}
\clearpage

\section{The RG Flow with a tachyon}

We now consider the RG flow itself, rather than the fixed points. 
When the tachyon is zero, the beta functions were computed in \cite{beta,cfmp}.
When there is a non-zero tachyon, the beta functions computed in \cite{tachyon}
yield the flow equations (assuming the central charge term in the dilaton
beta function is zero)

\bea
\dot g_{ij}&=&-\alpha'(R_{ij}+2\phi_{|ij}-T_{,i}T_{,j});\\
\dot\phi&=&{\alpha'\over
2}(\Delta\phi-2|\nabla\phi|^2 - V(T) ));\\
\dot T&=&\alpha'(\Delta T-2\nabla\phi\cdot\nabla T - V'(T)).
\label{tachflow}
\eea
 
The dot refers to derivative with respect to the RG flow parameter (logarithm of
scale).

An interesting feature of these flow equations seems to be that the dilaton 
cannot be decoupled from the metric and tachyon flow by a choice of gauge
(as it can be when the tachyon is zero). Curiously, if the beta function 
of the tachyon were multiplied by an overall factor of ${1 \over 2}$, then
a decoupling of the dilaton would have been possible in the resulting
flow equations. 

We now look for special solutions to the flow equations. We are 
motivated by the conjecture that in certain situations, the RG flow can 
model on-shell time evolution (for an elaboration of when this applies,
see \cite{HMT}). By on-shell time evolution, we mean 
spacetime metrics obtained by solving the fixed point equations. 

We are interested in whether there are RG flow
solutions that are similar to the tachyon cosmologies discussed by 
Yang and Zweibach \cite{YZ}.

The tachyon cosmology of Yang and Zweibach is a solution of 
Eq.(\ref{eq1-1}-\ref{eq1-3})
with $c=0$ and a Lorentzian metric of the form
\be
ds^2=-dt^2+a(t)\delta_{ij}dx^i dx^j,
\ee
and with dilaton and tachyon fields dependent only on the time coordinate $t$.  
In \cite{YZ}, tachyon-induced rolling is studied. When the tachyon
potential is of the form Eq.(\ref{tachpot}) (i.e, only leading order term), 
then tachyon-induced rolling 
implies that $da/dt = 0$, so that the string frame metric does not
evolve. For other potentials, it is possible to find
solutions with $da/dt = 0$.

We examine RG flow equations for two tachyon potentials studied in \cite{YZ}. 
In an
abuse of notation, in what follows, $t$ will be used to denote the RG flow
parameter.
This makes it easier to compare with the solutions in \cite{YZ}; however,
the RG flow parameter is not, in general, the same as the dynamical time
of those solutions. Rather, we would like to explore, for two choices
of potentials $V(T)$, if $t$ indeed behaves like the dynamical time. 
We look for solutions to RG flow with {\em Riemannian metrics} and the 
ansatz that the tachyon and the dilaton only depend on $t$ (RG flow parameter) 
with the metric
of the form
\be
ds^2=a(t)\delta_{ij}dx^i dx^j .
\label{ansatz}
\ee

With this ansatz, it trivially follows that $a(t)$ is a constant; so 
the metric does not evolve for any choice of tachyon potential.

Now we assume a tachyon potential of the form Eq.(\ref{tachpot}) with just the lowest
order term.
We get the solutions 
\bea
a(t)&=&a ;\none  
T(t)&=&T_0e^{\alpha'm^2 t}; \none 
\phi(t)&=&{T_0^2 \over 8}e^{2\alpha'm^2 t}.
\eea
Here, $a$ is used to denote a constant.
So, in agreement with Yang and Zweibach, the metric (if interpreted
as a string frame metric) is 
constant, the tachyon induces rolling at $t = - \infty$ and the 
solutions themselves look very similar near $t = -\infty$ to the 
solutions in \cite{YZ} for the corresponding potential, except that
$m$ is now replaced by $m^2$. As $t \to \infty $, both the tachyon and 
dilaton diverge.
If the above metric were indeed a string frame metric, the `Einstein
frame metric' 
\be
g^E_{ij}:=e^{-2\phi(t)}g_{ij},
\ee
is constant (in $t$) and regular near 
$t=-\infty$.  On the other hand, as $t\to+\infty$, the
Einstein frame metric crunches to a point (conformal factor goes to 
zero).

This nice similarity to solutions of dynamical evolution does not 
hold for other choices of potentials. 
For $V(T) = - {1 \over 2}~m^2  \left ( T^2 -  T^{4}/4 \right )$,
there are two branches of solutions to the RG flow equations. One branch
corresponds to singular tachyonic initial condition, so we disregard it.
For the other branch, we find that 
\bea
a(t)&=&a ;\none
T(t)&=& \sqrt{2} 
\frac{T_{0} e^{\alpha' m^2 t}}{\sqrt{1 + T_{0}^{2} e^{ 2 \alpha' m^2 t}}};\none
\phi(t)&=& \phi_{0} + \alpha' {m^2 \over 4} t - {1 \over 8} 
\ln \left [ \frac{T_{0}^{2} e^{2 \alpha' m^2 t}}{1 + T_{0}^{2}e^{2 \alpha' m^2 t}} \right ] -  \frac{1}{8(1 + T_{0}^{2}e^{2 \alpha' m^2 t})}.
\eea

As we said earlier, the metric does not evolve.
The behaviour of the tachyon and the dilaton for the above solution in 
various limits is as follows:
\begin{itemize}
\item
Although the dilaton has a linear term, one finds on careful analysis that 
as $t \to - \infty$, the linear term cancels out with contributions
coming from other terms. The integration constant $\phi_0$ can be chosen
such that the dilaton goes to zero as $t \to - \infty$. In fact, it
then goes to zero as $e^{2 \alpha' m^2 t}$. In this limit, the tachyon goes
to zero slower, as $e^{ \alpha' m^2 t}$. Therefore the tachyon induces the
rolling at $t = - \infty$.
\item
As $t \to \infty$, the dilaton grows linearly with $t$. The 
tachyon instead settles to the constant value $\sqrt{2} T_0 $.
Thus the corresponding `Einstein frame' metric crunches in this limit to
a point, due to the behaviour of the dilaton.

\end{itemize}
This solution (particularly the tachyon) behaves differently from Yang and Zwiebach's dynamical
solution for the potential we chose, except at early times when the tachyon induces the rolling. This is not in itself surprising. The
phase space of solutions to the dynamical equations is bigger than that for
the RG flow, and the similarity between the two seems to hold only for a
particular time dependence of the tachyon and dilaton. 

\bigskip
\noindent
{\bf Acknowledgments}\\
The authors wish to thank the Natural Sciences and Engineering Research 
Council of Canada for partial support. V.S would like to thank
B. Sathiapalan for a discussion on the tachyon beta function and E. Woolgar
for comments on the paper. We would also like to thank the referee for
bringing \cite{peet} to our attention.

\end{document}